\documentclass[sn-basic]{sn-jnl}
\usepackage{lmodern}
\usepackage{subcaption}
\usepackage{tcolorbox}
\usepackage{adjustbox}
\usepackage{graphicx}%
\usepackage{multirow}%
\usepackage{amsmath,amssymb,amsfonts}%
\usepackage{amsthm}%
\usepackage{mathrsfs}%
\usepackage[title]{appendix}%
\usepackage{soul, xcolor}%
\usepackage{textcomp}%
\usepackage{manyfoot}%
\usepackage{booktabs}%
\usepackage{url}%
\usepackage{longtable}
\usepackage[ruled,linesnumbered]{algorithm2e}
\usepackage{listings}%
\SetKwInput{KwInput}{Input}                
\SetKwInput{KwOutput}{Output} 

\begin{document}

\title[QSVC Vs. SVC for Software Bug Prediction]{Comparative Analysis of Quantum and Classical Support Vector Classifiers for Software Bug Prediction: An Exploratory Study}


\author*[1]{\fnm{Md} \sur{Nadim}}\email{mnadims.cse@gmail.com}

\author[2]{\fnm{Mohammad} \sur{Hassan}}\email{mohammadhassan@upei.ca}

\author[1]{\fnm{Ashis} \sur{Kumar Mandal}}\email{ashis62@gmail.com}
\author[1]{\fnm{Chanchal} \sur{K. Roy}}\email{chanchal.roy@usask.ca}
\author[1]{\fnm{Banani} \sur{Roy}}\email{banani.roy@usask.ca}
\author[1]{\fnm{Kevin} \sur{A. Schneider}}\email{kevin.schneider@usask.ca}

\affil*[1]{\orgname{University of Saskatchewan}, \city{Saskatoon}, \state{SK}, \country{Canada}}

\affil[2]{\orgname{University of Prince Edward Island}, \city{Charlottetown}, \state{PE}, \country{Canada}}



\abstract{\textbf{Purpose:} Quantum computing promises to transform problem-solving across various domains with rapid and practical solutions. Within Software Evolution and Maintenance, Quantum Machine Learning (QML) remains mostly an underexplored domain, particularly in addressing challenges such as detecting buggy software commits from code repositories.

\textbf{Methods:} In this study, we investigate the practical application of Quantum Support Vector Classifiers (QSVC) for detecting buggy software commits across 14 open-source software projects with diverse dataset sizes encompassing 30,924 data instances. We compare the QML algorithm PQSVC (Pegasos QSVC) and QSVC against the classical Support Vector Classifier (SVC).  Our technique addresses large datasets in QSVC algorithms by dividing them into smaller subsets. We propose and evaluate an aggregation method to combine predictions from these models to detect the entire test dataset. We also introduce an incremental testing methodology to overcome the difficulties of quantum feature mapping during the testing approach.

\textbf{Results:} The study shows the effectiveness of QSVC and PQSVC in detecting buggy software commits. The aggregation technique successfully combines predictions from smaller data subsets, enhancing the overall detection accuracy for the entire test dataset. The incremental testing methodology effectively manages the challenges associated with quantum feature mapping during the testing process.

\textbf{Conclusion:} We contribute to the advancement of QML algorithms in defect prediction, unveiling the potential for further research in this domain. The specific scenario of the Short-Term Activity Frame (STAF) highlights the early detection of buggy software commits during the initial developmental phases of software systems, particularly when dataset sizes remain insufficient to train machine learning models.
}




\maketitle

\section{Introduction}
\label{introduction}
The emergence of quantum computing offers a new kind of computing that could change how we solve problems [\cite{hidary2019quantum}]. It opens lots of exciting possibilities in different areas. The potential of quantum computing for rapid and effective solutions has sparked considerable interest among researchers seeking innovative approaches to address complex computational problems [\cite{khrennikov2021roots}]. Use of Quantum technology [\cite{Schuld2017SimpleQ}] and Artificial Intelligence (AI), giving rise to an expanding research field known as Quantum Machine Learning (QML) [\cite{QMLPHYSE}]. QML algorithms represent a fusion of quantum principles with classical machine learning techniques [\cite{QML1, QML2, QMLTutorial}], leveraging advancements in quantum technologies and employing quantum software engineering methodologies to enhance traditional Machine Learning (ML) algorithms for classical datasets. One of the biggest challenges to employing QML and other Quantum Computing [\cite{QuantumComputing, Garcia2022SystematicLR}] methods on classical data is to encode that classical data to quantum space effectively and avoid the associated exponential complexity of that encoding process [\cite{QuantumEnhancedFS}].

The quest for efficient and reliable methods of detecting software bugs remains a critical challenge [\cite{Rel:2015:Yang:Deep:JIT:Defect:Prediction}] in the software engineering research domain. While quantum computing [\cite{QuantumComputing, Garcia2022SystematicLR}] holds immense promise, its widespread availability and practical application in real-world scenarios are still in the preliminary stages. Many quantum algorithms are under active development, and numerous potential application areas are yet to be fully explored. Quantum Machine Learning (QML) remains one of these unexplored research territories in Software Engineering (SE). Specifically, applying QML techniques for detecting buggy software commits from source code repositories remains a promising but challenging endeavor due to the complex interplay between quantum principles and traditional software development practices. The accessibility of real quantum computers remains limited, prompting researchers to rely on quantum simulators and quantum-inspired classical algorithms to explore the problem-solving capabilities of quantum computing across diverse domains. However, these alternatives present challenges, including managing extremely high runtimes for processing large datasets. 

In this study, we embark on a journey to investigate the practical application of Quantum Support Vector Classifiers (QSVC) for detecting buggy software commits. Our research focuses on a comprehensive analysis of QSVC performance across a diverse range of open-source software projects, comprising a substantial dataset encompassing 30,924 data instances listed in Table \ref{tab:dataset-summary}. These datasets came from 14 software projects frequently employed in software bug detection studies [\cite{Menzies2004AssessingPO, BugMobileApp, Kamei:2013:LES:2498737.2498844, ApacheJIT, JITOS, NadimSCG, NadimBIC}].  These datasets contain identified buggy and non-buggy software commits through automatic data labeling processes. These automatic processes follow some rules to determine a software commit as buggy or non-buggy; for example, a commit can be labeled buggy if followed by another commit that fixes a bug introduced by the first commit [\cite{NadimBIC}]. The SZZ algorithm [\cite{2019:Borg:SZZUnleashed}] is well-known for performing automatic identification of bug-inducing commits from software projects. There are some other studies [\cite{2019:FSE:Wen:EEC:3338906.3338962, Quach2021, Quach2021empirical}] which performed manual verification of the buggy commits identified by the automatic process and published their verified datasets, but these datasets usually contain a very limited number of data instances, which may not be sufficient for training machine learning models. In this investigation, we selected all the datasets shown in Table \ref{tab:dataset-summary} labeled automatically by these previous studies.  

To ensure a comprehensive analysis, we randomly sampled 14 datasets with varying total commit instances, ranging from 498 to 8604, with an average of 2209 instances across each subject system. We believe that this diverse selection of dataset sizes contributes significantly to the generalizability of the findings of this investigation. Across these datasets, the number of training and testing samples varies, from 348 to 6883 for training, 35 to 688 for tuning, and 150 to 1721 for testing, resulting in 23,968 training samples, 2,397 tuning samples, and 6,956 testing data samples. Utilizing such a diverse array of data samples in our experiments infuses confidence in the robustness of our findings and enables us to address the research question at hand effectively. To contextualize our findings, we compare the performance of two variations of the QSVC algorithm available as IBM Qiskit [\cite{Qiskit}] Python library against the classical Support Vector Classifier (SVC) from the widely-used Scikit-learn [\cite{scikit-learn}] Python library.

We evaluated our results to answer the following research questions (\textbf{RQs}).

\begin{itemize}
    \item \textbf{RQ1:} How does the Quantum SVC algorithm perform in Short-term Activity Frames (STAF) compared to the traditional SVC algorithm in buggy software commit detection?

    Data scarcity is a common problem in software bug detection datasets, especially when a software project is relatively new in the early stage of any newer launch/version and does not have much historical data to train a detection model. To address this RQ, we compare the performance of Quantum SVC and Classical SVC with smaller chunks of the training dataset.

    \item \textbf{RQ2:} Can we apply Quantum SVC algorithms to a large dataset of real-life software bug detection problems? 
    
    To tackle this research question, our study observes the training and testing durations of both QSVC and PQSVC algorithms as the dataset's instance size increases incrementally. We thoroughly logged the runtime requirements for training and testing these algorithms across a spectrum of data instance quantities and presented our findings and an in-depth discussion in the results section.

    \item \textbf{RQ3:} Does aggregation of trained QSVC models on smaller chunks of datasets make a better globally trained QSVC model to deal with large datasets? 
    
    When addressing RQ2, we encountered significant challenges in training QSVC with datasets containing more than 500 instances, primarily due to its sluggish runtime. QSVC's performance deteriorates notably with larger datasets, rendering it unresponsive and failing to yield any output. To mitigate this issue, we introduce a novel approach wherein we train multiple smaller models using subsets of the data, each comprising 500 instances. Subsequently, we aggregate the predictions from these smaller models to derive the overall detection of the global QSVC model. This strategy was empirically validated across six subject systems, each characterized by datasets ranging from 93 to 688 instances, and tested on datasets ranging from 400 to 1721 instances.
\end{itemize}

\begin{table}[ht]
\centering
\caption{Dataset Summary in Decreasing Order of Size}
\label{tab:dataset-summary}
\begin{tabular}{clcccc}
\hline
\multicolumn{1}{|c|}{\textbf{\begin{tabular}[c]{@{}c@{}}SL.\\ No.\end{tabular}}} & \multicolumn{1}{c|}{\textbf{\begin{tabular}[c]{@{}c@{}}Subject\\ System\end{tabular}}} & \multicolumn{1}{c|}{\textbf{\begin{tabular}[c]{@{}c@{}}Train\\ Instances\end{tabular}}} & \multicolumn{1}{c|}{\textbf{\begin{tabular}[c]{@{}c@{}}Tuning\\ Instances\end{tabular}}} & \multicolumn{1}{c|}{\textbf{\begin{tabular}[c]{@{}c@{}}Test\\ Instances\end{tabular}}} & \multicolumn{1}{c|}{\textbf{\begin{tabular}[c]{@{}c@{}}Dataset\\ Size\end{tabular}}} \\ \hline \hline
\multicolumn{1}{|c|}{1}                                                          & \multicolumn{1}{l|}{AnySoftK}                                                   & \multicolumn{1}{c|}{6883} & \multicolumn{1}{c|}{688}                                                                & \multicolumn{1}{c|}{1721}                                                            & \multicolumn{1}{c|}{8604}                                                             \\ \hline
\multicolumn{1}{|c|}{2}                                                          & \multicolumn{1}{l|}{Kiwis}                                                             & \multicolumn{1}{c|}{4905}    & \multicolumn{1}{c|}{491}                                                                 & \multicolumn{1}{c|}{1227}                                                            & \multicolumn{1}{c|}{6132}                                                             \\ \hline
\multicolumn{1}{|c|}{3}                                                          & \multicolumn{1}{l|}{Facebook}                                                          & \multicolumn{1}{c|}{3523}     & \multicolumn{1}{c|}{352}                                                                & \multicolumn{1}{c|}{881}                                                             & \multicolumn{1}{c|}{4404}                                                             \\ \hline
\multicolumn{1}{|c|}{4}                                                          & \multicolumn{1}{l|}{Jm1}                                                               & \multicolumn{1}{c|}{3369}     & \multicolumn{1}{c|}{337}                                                                & \multicolumn{1}{c|}{843}                                                             & \multicolumn{1}{c|}{4212}                                                             \\ \hline
\multicolumn{1}{|c|}{5}                                                          & \multicolumn{1}{l|}{OpenStack}                                                                & \multicolumn{1}{c|}{936}      & \multicolumn{1}{c|}{94}                                                                & \multicolumn{1}{c|}{404}                                                             & \multicolumn{1}{c|}{1340}                                                             \\ \hline
\multicolumn{1}{|c|}{6}                                                          & \multicolumn{1}{l|}{Camel}                                                             & \multicolumn{1}{c|}{928}    & \multicolumn{1}{c|}{93}                                                                  & \multicolumn{1}{c|}{400}                                                             & \multicolumn{1}{c|}{1328}                                                             \\ \hline
\multicolumn{1}{|c|}{7}                                                          & \multicolumn{1}{l|}{Jackrabbit}                                                        & \multicolumn{1}{c|}{556}    & \multicolumn{1}{c|}{56}                                                                  & \multicolumn{1}{c|}{240}                                                             & \multicolumn{1}{c|}{796}                                                              \\ \hline
\multicolumn{1}{|c|}{8}                                                          & \multicolumn{1}{l|}{QT}                                                                & \multicolumn{1}{c|}{472}  & \multicolumn{1}{c|}{47}                                                                    & \multicolumn{1}{c|}{204}                                                             & \multicolumn{1}{c|}{676}                                                              \\ \hline
\multicolumn{1}{|c|}{9}                                                          & \multicolumn{1}{l|}{Bitcoin}                                                           & \multicolumn{1}{c|}{460}   & \multicolumn{1}{c|}{46}                                                                   & \multicolumn{1}{c|}{200}                                                             & \multicolumn{1}{c|}{660}                                                              \\ \hline
\multicolumn{1}{|c|}{10}                                                         & \multicolumn{1}{l|}{Tomcat}                                                            & \multicolumn{1}{c|}{452}   & \multicolumn{1}{c|}{45}                                                                   & \multicolumn{1}{c|}{194}                                                             & \multicolumn{1}{c|}{646}                                                              \\ \hline
\multicolumn{1}{|c|}{11}                                                         & \multicolumn{1}{l|}{Ambari}                                                            & \multicolumn{1}{c|}{410}      & \multicolumn{1}{c|}{41}                                                                & \multicolumn{1}{c|}{178}                                                             & \multicolumn{1}{c|}{588}                                                              \\ \hline
\multicolumn{1}{|c|}{12}                                                         & \multicolumn{1}{l|}{Mongo}                                                             & \multicolumn{1}{c|}{368}       & \multicolumn{1}{c|}{37}                                                               & \multicolumn{1}{c|}{158}                                                             & \multicolumn{1}{c|}{526}                                                              \\ \hline
\multicolumn{1}{|c|}{13}                                                         & \multicolumn{1}{l|}{Oozie}                                                             & \multicolumn{1}{c|}{358}      & \multicolumn{1}{c|}{36}                                                                & \multicolumn{1}{c|}{156}                                                             & \multicolumn{1}{c|}{514}                                                              \\ \hline
\multicolumn{1}{|c|}{14}                                                         & \multicolumn{1}{l|}{Lucene}                                                            & \multicolumn{1}{c|}{348}          & \multicolumn{1}{c|}{35}                                                            & \multicolumn{1}{c|}{150}                                                             & \multicolumn{1}{c|}{498}                                                              \\ \hline 
                                                                                 &                                                                                        &                                                                              23,968 & 2,397
         &                                                                             6,956
         & 30,924                                                                                
 \end{tabular} 
\end{table}

We organize the subsequent sections of this paper by describing the background of this study in section \ref{background}, methodology in section \ref{methodology}, result and performance comparison of our study in section \ref{result-discussion}, some threats to the validity in section \ref{threats-validity}, related works in section \ref{related-work}, and finally we conclude the study mentioning some future directions in section \ref{conclusion-future}.

\section{Background}
\label{background}
\textbf{Quantum computing} is a revolutionary approach to computation that employs the principles of quantum mechanics to solve complex problems that are beyond the capabilities of classical computers [\cite{hidary2019quantum}]. At the heart of quantum computing is the qubit, which is a unit that can exist in a superposition of classical zero and one state, providing unparalleled computational versatility compared to classical bits. Quantum computers are adept at tackling complex challenges [\cite{khrennikov2021roots}] by utilizing quantum phenomena such as superposition, entanglement, and interference. Empirical and theoretical research highlights their ability to excel in machine learning, optimization, and simulations, surpassing classical counterparts in efficiency [\cite{hellstem2021hybrid, cho2021quantum}].

\textbf{Quantum Machine Learning (QML)} offers a promising approach whereby quantum information processing is harnessed for various machine learning tasks, including clustering, regression, and classification [\cite{QMLQiskitPY, QMLPHYSE, biamonte2017quantum, QSVC}]. QML leverages the extraordinary capabilities of quantum systems to tackle complex problems that conventional computers struggle to solve. Quantum-enhanced machine learning [\cite{dunjko2016quantum}] has the potential to advance the fields of supervised, unsupervised, and reinforcement learning [\cite{lamata2017basic, dong2008quantum, Qsupervised}], offering quadratic efficiency improvements and exponential performance gains over limited periods in various learning scenarios. QML demonstrates superior capabilities in handling high-dimensional data and uncovering intricate patterns compared to Classical Machine Learning (CML) algorithms [\cite{ramezani2020machine}], particularly in specific application domains such as Grover’s Algorithm [\cite{GroverAlgo}]. This study aims to investigate and compare the Quantum Support Vector Classifier (QSVC) performance against its classical counterpart using six realistic datasets for detecting buggy software commits.



\textbf{The Quantum Support Vector Classifier (QSVC)} is a quantum machine learning (QML) algorithm specifically tailored for binary classification, drawing from the principles of Support Vector Machines (SVM). Numerous QSVC algorithms have been proposed to date [\cite{ramezani2020machine}]. The fundamental disparity between quantum and classical SVC lies in kernel computation. In classical SVC, the kernel function is typically predetermined, whereas in quantum computing, it is derived through the utilization of a quantum circuit [\cite{Heredge2021}]. Within the QSVC framework, input data undergoes encoding into a quantum state, which is subsequently mapped onto a high-dimensional quantum feature space. The advancement of quantum machine learning relies heavily on distinctive quantum feature maps, including the Z-feature map, the ZZ-feature map, and the Pauli-feature map [\cite{QuantumEnhancedFS}]. These feature maps play a crucial role in driving progress in quantum machine learning methodologies. After careful examination, \citet{QMLIEEEAccess} determined the Z-feature map to be the most effective approach among various explored options. Therefore, we have employed this feature mapping technique in our investigation.


\textbf{The Qiskit Machine Learning} library in the Python programming language [\cite{QMLQiskitPY, Qiskit}] stands out as a potent instrument for easier practical application of quantum machine learning algorithms. It delivers a spectrum of tools, including Quantum Kernels, Quantum Neural Networks (QNNs), and an array of learning algorithms, offering users a robust framework to explore the boundaries of quantum-assisted machine learning. 

\begin{figure}
\centering
\includegraphics[width=\textwidth] {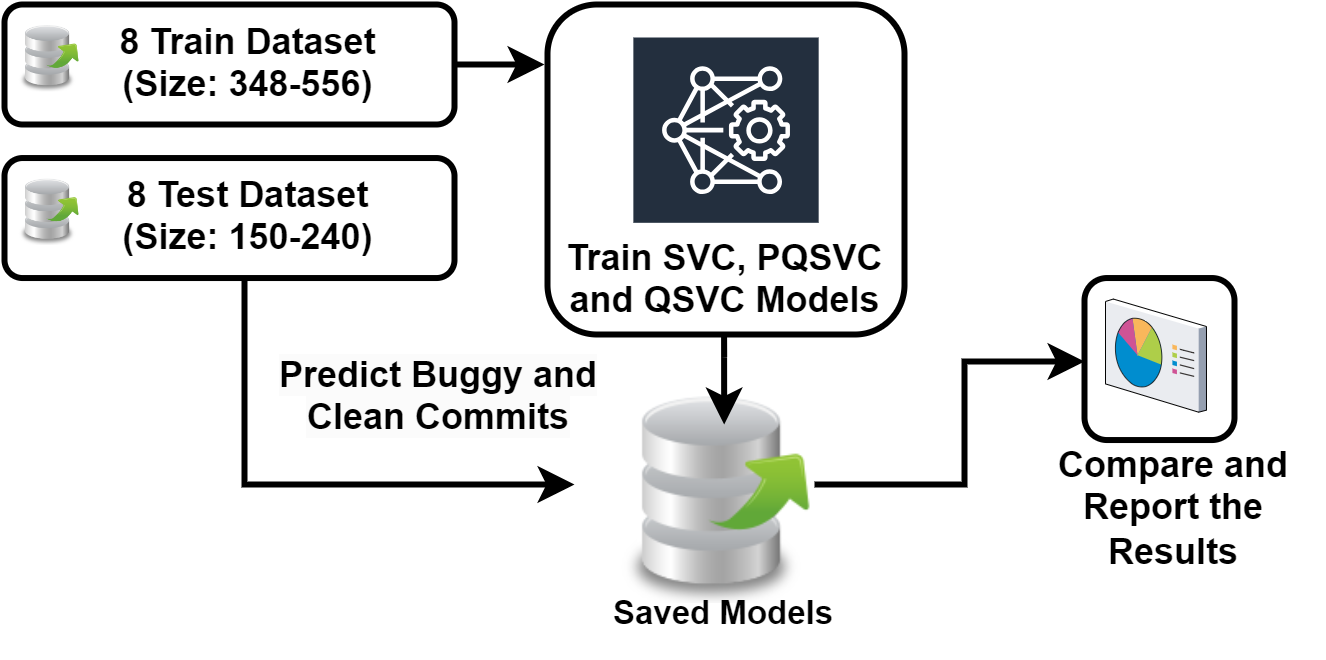}
\caption{Short-term Activity Frame (STAF)}
\label{fig:staf}
\end{figure}

\section{Methodology}
\label{methodology}
The QISKIT library offers variations of the Quantum SVC algorithm, specifically PQSVC and QSVC, designed for the Python programming language [\cite{Qiskit}]. Analogous to the classical SVC implementation found in the Scikit-learn Python library [\cite{scikit-learn}], these quantum algorithms can be applied to datasets for classification tasks. However, a significant challenge lies in effectively leveraging the Quantum SVC algorithms with varying numbers of training and testing instances to achieve reliable classification results within a realistic runtime. Our research addresses this challenge by conducting experiments on 14 datasets containing instances of both buggy and clean software commits, with instances ranging from 498 to 8604. Table \ref{tab:dataset-summary} summarizes the train and test instances used for classification. Notably, with smaller datasets containing fewer than 500 instances, all SVC, PQSVC, and QSVC algorithms produce results within a realistic runtime. We perform our investigation using the following key steps.

\subsection{Dataset Preparation}
We categorized our dataset into two groups based on the number of data instances they contained. The first category comprised datasets with a larger number of instances, ranging from 928 to 6883. The second category consisted of eight datasets with data instances ranging from 348 to 556. Before delving into our investigation, we carefully segregated training and testing data instances from all 14 subject systems. This careful separation of testing data guarantees that they remain unseen to the trained models, ensuring the integrity of our evaluation. Leveraging these distinct sets of training and testing data, we evaluated the performance of our models, storing them on disk for precise effectiveness assessment. 

\subsection{Verifying the Short Term Activity Frame (STAF)}
We illustrate the procedural steps and the number of training and testing data instances of this phase of our investigation in Figure \ref{fig:staf}. During this stage, our objective was to validate the specific suitability of Quantum Support Vector Classifiers (QSVC) for analyzing the Short-Term Activity Frame (STAF) of a software system. This phase captures the early stages of a software project, where the dataset size available for training and testing machine learning models remains limited.

In this investigation, we worked with eight datasets, each containing a relatively smaller number of instances. Given the reduced size of these datasets, we opted to train Classical Support Vector Classifiers (SVC), QSVC, and PQSVC algorithms without partitioning the training data into smaller subsets. Subsequently, we stored the trained models on disk for later evaluation using testing datasets. Our aim in testing these eight subject systems was to reveal the performance of Quantum Support Vector Classifiers (QSVC) when trained on datasets of limited size — a typical scenario, particularly for developing software systems or newer versions of existing software.

We then conducted a comparative analysis between the performance of QSVC and two other classifiers: Classical Support Vector Classifiers (SVC) and Pegasos QSVC, which represents an alternative implementation of the QSVC algorithm.

\begin{figure}
\centering
\includegraphics[width=\textwidth] {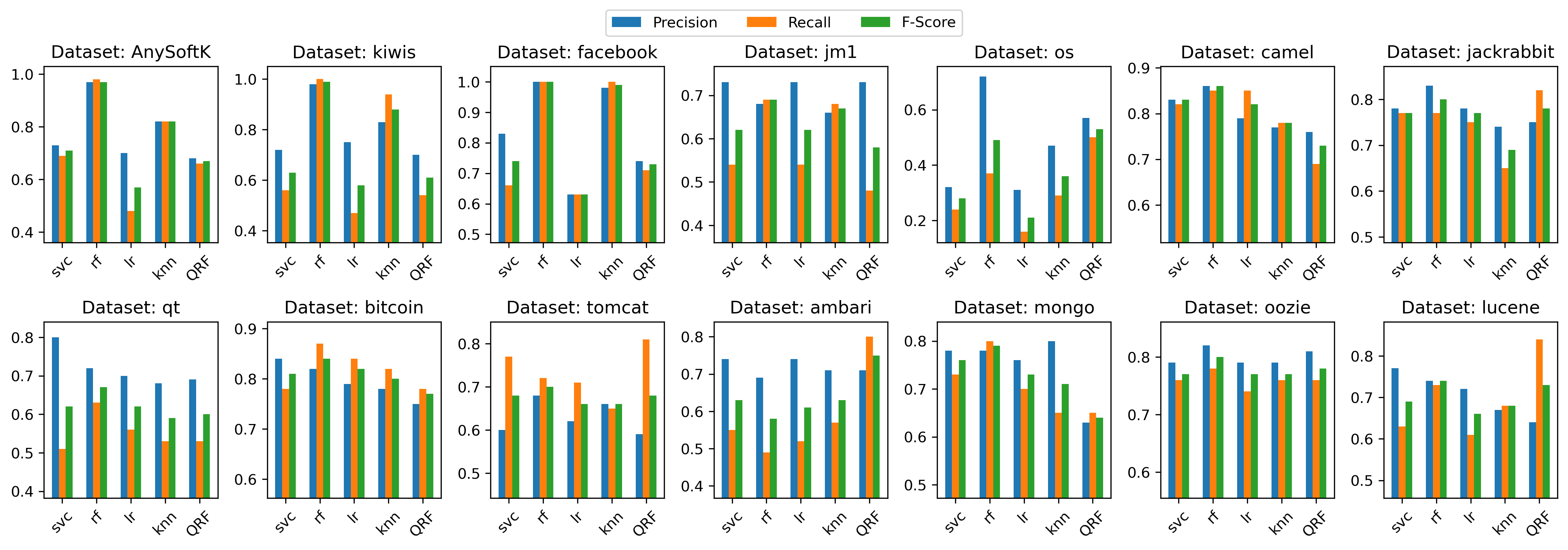}
\caption{Comparison of Precision, Recall, F1~Score of Quantum Random Forest (QRF) with popular classical ML (CML) algorithms. The datasets are in decreasing order as shown in the Table \ref{tab:dataset-summary}}
\label{fig:preliminaryStudy}
\end{figure}

\subsection{Train QSVC with Large Number of Dataset Instances}
We perform two preliminary studies to evaluate, whether we can run QML algorithms utilizing a large number of training and testing data instances which we can use easily using Classical ML (CML) algorithms. We use QSVC and Quantum Random Forest (QRF) algorithms to predict buggy commit instances from all the 14 datasets used in this investigation. QRF algorithm is proposed by \citet{Srikumar2024}, which uses a Support Vector Machine (SVM) [\cite{SVMbook}] and Quantum Kernel Estimation (QKE) [\cite{QKE2019}] approach to form individual decision trees as the unit of the Random Forest algorithm [\cite{RandomForestAlgo}]. \citet{Srikumar2024} applied their QRF algorithm on randomly sampled 300 data instances from each dataset where they used 180 instances for training and 120 instances for testing the dataset. The QRF algorithm constructs different decision trees by using various segments of the training dataset. They applied their proof of concept on such a limited number of data instances to manage computational complexity effectively. However, using such a limited number of data instances does not fully reflect the demands and challenges of practical bug prediction scenarios, where larger datasets are typically necessary for robust model training.

Data instances in 14 subject systems used in our investigation shown in Table \ref{tab:dataset-summary} range from 498 (Lucene) to 8604 (AnySoftKeybord), totaling 30,924 instances. We attempted to apply both Quantum Random Forest (QRF) and Quantum Support Vector Classifier (QSVC)  algorithms on the entire data instances from all subject systems. In our preliminary investigation, we found that QRF can complete training and testing on all the data instances, but the time taken by this algorithm ranges from 7.66 minutes to 88.82 minutes in different subject systems. Despite this extended runtime, QRF did not achieve better predictive performance for buggy commits compared to popular classical machine learning (CML) algorithms, such as Support Vector Classifier (SVC), Random Forest (RF), Logistic Regression (LR), and K-Nearest Neighbors (KNN). We presented that comparison in Figure \ref{fig:preliminaryStudy}, where we can notice, that the Precision, Recall, and F1~Score of QRF are consistently lower than the CML algorithms in most of the datasets. QRF is performing slightly better compared to the CML algorithms only in 2 (Jackrabbit and Ambari) out of 14 datasets. This outcome suggests that, although QRF can handle large datasets and requires considerable processing time, it generally fails to outperform CML algorithms. CML algorithms achieve consistently better performance than QRF in most cases. When applying the QSVC algorithm to all the datasets in our investigation, we observed that it often became unresponsive, resulting in extended processing times without generating any output. Through trial and error, we observed that the QSVC algorithm successfully produced results only when the number of data instances per dataset was below 500. For datasets with larger numbers of instances, the algorithm failed to complete its execution, even after running continuously for several days.

These preliminary findings motivated us to develop an innovative approach for applying Quantum Machine Learning (QML) algorithms to predict software bugs in larger datasets. Given that both Quantum Support Vector Classifier (QSVC) and Quantum Random Forest (QRF) rely on SVM as their underlying model, we proposed a chunk-based training and testing method for the QSVC algorithm. This approach enables the QSVC algorithm to process larger data instances by dividing datasets into smaller segments train multiple models and save them to be used during the testing process.

\begin{figure}
\centering
\includegraphics[width=\textwidth] {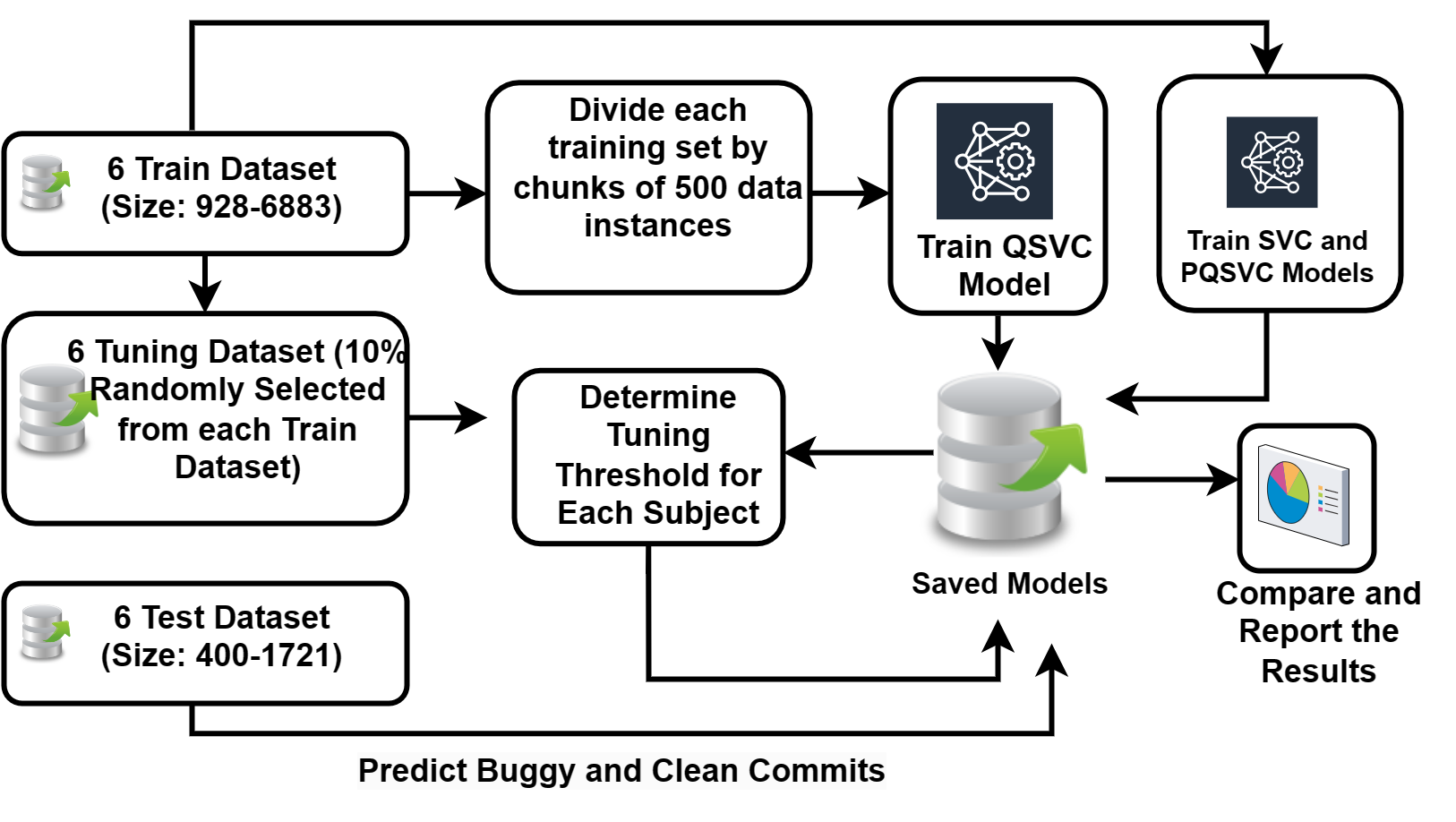}
\caption{Aggregation and Tuning Strategy of Chunk Models to Perform Global QSVC Prediction}
\label{fig:chunkAggregate}
\end{figure}

Figure \ref{fig:chunkAggregate} presents the method and the number of employed dataset instances of applying the QSVC algorithm on larger datasets to predict buggy software commits. In this phase, we can not use a conventional approach with the whole training and testing data at once due to the exponential complexity of the Qunatum Feature Mapping [\cite{QuantumEnhancedFS}] to train and test the QSVC algorithm. We proposed an updated methodology that can iteratively handle a larger dataset while using this algorithm. 

In this phase of our investigation, we chose to partition the training data from the six subject systems into smaller chunks, each containing 500 data instances. To elaborate further, given the dataset size for AnySoftK (6883 instances), this resulted in 13 data chunks containing 500 data instances and one chunk containing the rest of the data. Subsequently, we trained a QSVC model with each of these 14 data chunks, saving the corresponding trained models onto disk. The testing data sets were pre-allocated and stored on the disk as well. Following this, we evaluated each trained model using the respective data chunks from our test dataset, yielding buggy commit predictive outcomes for the testing dataset for each training chunk.

\begin{figure}
\centering
\includegraphics[width=0.75\textwidth] {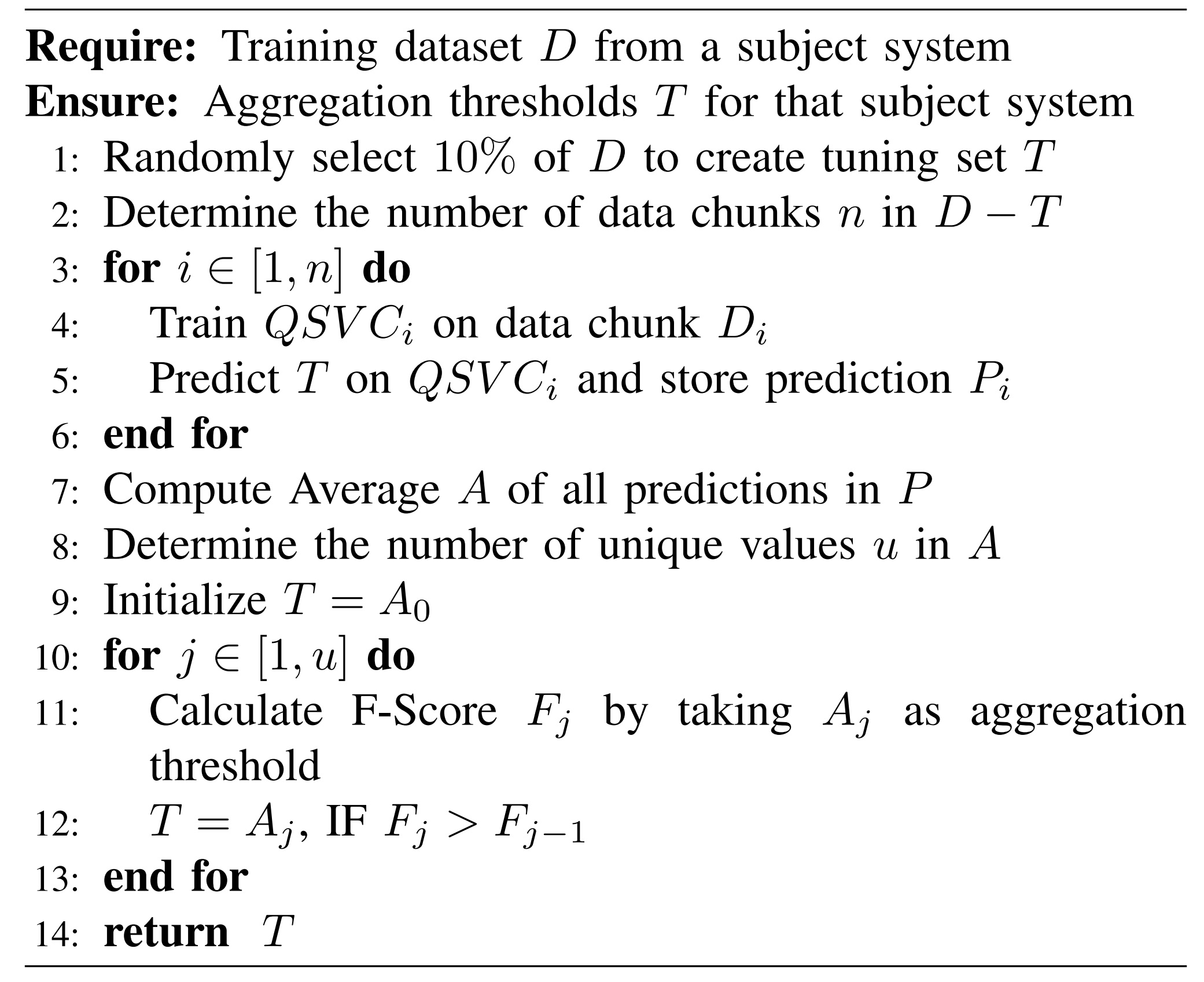}
\caption{Calculating Aggregation Threshold for Maximizing F-Score}
\label{alg:thresholdcalc}
\end{figure}

\subsection{Determining the Aggregation Threshold}
To obtain an aggregated result for the testing dataset of each subject system, we implemented an aggregation strategy on the results obtained from each train chunk model. We apply the Homogeneous Ensemble [\cite{HomogeneousEnsemble}] approach, which involves combining multiple models (often called “weak learners”) of the same type to enhance overall predictive performance. Figure \ref{alg:thresholdcalc} shows the aggregation strategy implemented in this investigation. We first employ a simple averaging technique to aggregate the predictions from each of the n-trained models for each subject system. The aggregation technique involved computing the average predicted results for each data instance across all n trained models as shown in Equation \ref{eq:average-prediction}. Subsequently, based on the calculated average, we assigned a binary value \textbf{0} representing \textbf{clean} and \textbf{1} representing \textbf{buggy} to each software commit in our test dataset. In the conversion of the average value to its buggy or non-buggy equivalence, we utilize an aggregation threshold value. For example, let a test instance be detected as buggy and non-buggy by eight and six of the n=14 trained chunk models. Therefore, we can calculate the average prediction by $8/14 = 0.57$. If we perform simple majority voting, we can take the threshold 0.50, and any average equal of above 0.50 can converted to 1, meaning buggy test instance, and the other as 0, meaning non-buggy test instance. To avoid doing that simple majority voting, we perform a threshold tuning method by taking 10\% of the total training data instances. Using that threshold tuning method, we calculate the optimized threshold for each subject system, which can be either below or above the 0.50 value. We describe our tuning process in the subsequent paragraphs. 

\begin{equation}
\centering
\text{Average Prediction} = \frac{1}{n} \sum_{i=1}^{n} \text{Prediction}_i
\label{eq:average-prediction}
\end{equation}

\subsection{Tuning Chunk Models for Threshold Optimization}
\begin{figure}
    \centering
    \begin{subfigure}{\textwidth}
        \includegraphics[width=0.85\linewidth]{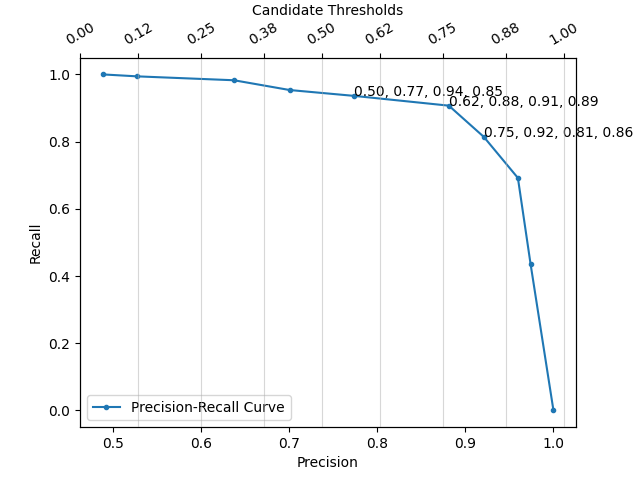}
        \caption{Facebook}
    \end{subfigure}    
    
    

     \caption{Utilizing Precision-Recall Curve to Determine Optimal Tuning Threshold for Training Aggregated Chunk-QSVC into Global QSVC. Each data point on the curve presents the associated threshold, precision, recall, and F-score values.}. 
    \label{fig:calcThreshold}
\end{figure}

\begin{figure}
\centering
\includegraphics[width=\textwidth] {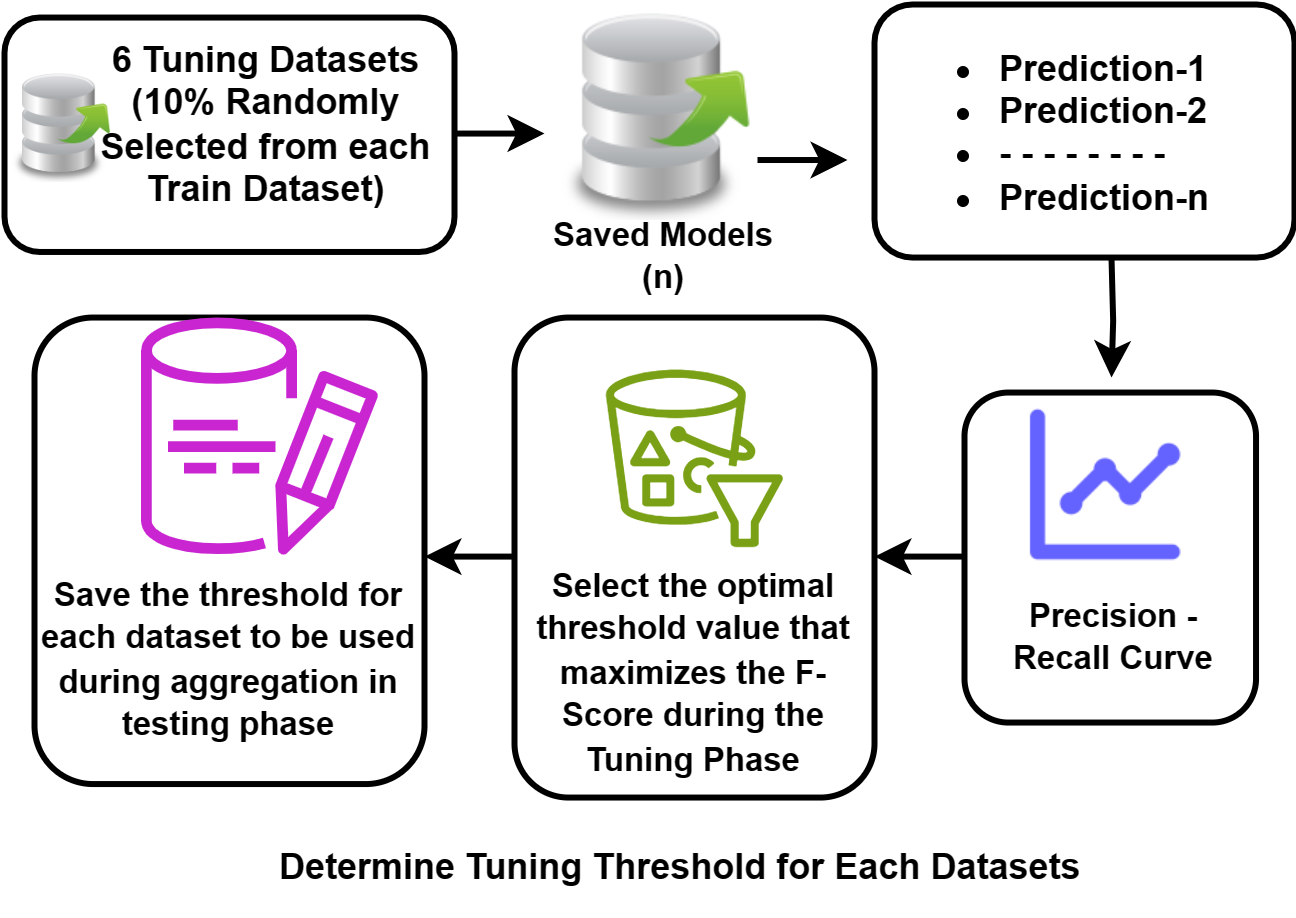}
\caption{Selecting Aggregation Threshold in Tuning Process}
\label{fig:tuningThreshold}
\end{figure}

Figure \ref{fig:tuningThreshold} demonstrates the methodology for determining the aggregation threshold value in our tuning process. This process leverages all the trained chunk models to evaluate the tuning chunk derived from 10\% of the entire training dataset. Additionally, we show the algorithm in Figure \ref{alg:thresholdcalc} for calculating aggregation threshold. The process is also demonstrated in  Figure \ref{fig:calcThreshold}. We conduct predictions on the tuning dataset using all trained models on data chunks, subsequently plotting a precision-recall curve to ascertain the threshold value that yields the highest F-score on the tuning datasets. This identified threshold value is then saved for later utilization in the testing phase. This process also offers an opportunity to prioritize any desired performance metric, such as precision, recall, or other relevant measures, depending on the specific bug prediction scenario. Our novel method for determining the threshold value grants practitioners greater flexibility to choose and optimize the metric that aligns with their goals. In our approach, we selected the F1~Score for optimization during the tuning process. The F1~Score, as the harmonic mean of precision and recall, provides a balanced evaluation by integrating both measures effectively. After determining the tuning threshold, we use the threshold value in the testing process using the test dataset. 

\subsection{Testing the Model Performance}
\begin{figure}[ht]
\centering
\includegraphics[width=0.75\textwidth] {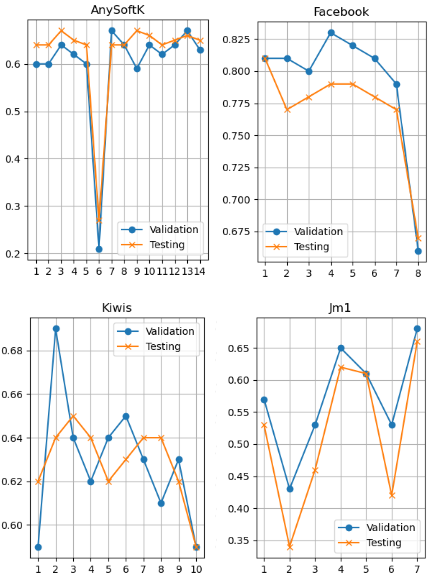}
\caption{Comparing F1~Score in Different Chunk Models of Four Datasets (AnySoftK, Facebook, Kiwis, Jm1). The X-axis represents different chunk models, and the Y-axis represents the corresponding F1~Score value in tuning and test datasets.}
\label{fig:validationVsTest}
\end{figure}

In Figure \ref{fig:validationVsTest}, we present a comparative analysis of F-scores between tuning and testing datasets across four subject systems characterized by a larger number of chunks within our study. Each scenario depicted in this figure exhibits a consistent pattern in F-score performance during both the tuning and testing phases. Notably, the F-scores observed in the testing phase consistently exhibit a slight decrement compared to those in the tuning phase. This discrepancy may be curbed by the fact that our tuning process entails randomly selecting a subset of the training dataset segments. Despite this marginal disparity, the primary similarity in F-score trends between the tuning and testing phases highlights the robust generalizability of the trained QSVC chunk models.

We replicated this methodology across all six datasets examined in our investigation, ensuring each dataset contained a sufficient number of data instances to generate at least two training data chunks. This aggregation process yielded a Global QSVC model results for each subject system, consolidating models trained on respective data chunks. Subsequently, we compared the performance of Classical SVC, PQSVC, and Global QSVC against each other to determine the efficacy of the global QSVC algorithm in comparison to the other algorithms. The results of our evaluation are presented and discussed in the Results and Discussion section of this study.

\pagebreak
\section{Result and Discussion}
\label{result-discussion}
In this section, we present the findings of our investigation and answer the subsequent research questions (\textbf{RQs}). 

\begin{tcolorbox}[colback=gray!10!white, colframe=gray!80!black, arc=4mm]
\textbf{RQ1:} How does the Quantum SVC algorithm perform in Short-term activity frames (STAF) compared to the traditional SVC algorithm in buggy software commit detection?
\end{tcolorbox}

To address \textbf{RQ1}, we present our findings in  Table \ref{tab:staf-result}, which illustrates the results obtained from our study when we use the Short-Term Activity Frames (STAF) working principle. We evaluate the performance of the classical support vector classifier (SVC) and two Quantum support vector classifiers (PQSVC and QSVC).  STAF refers to a common occurrence in software systems, typically observed during the early stages of the software development life cycle or in the context of new software versions with limited data instances. The scarcity of data instances in STAF scenarios often renders classical machine learning algorithms ineffective during training, leading to poor classifier performance. Therefore, our aim in this section is to assess the efficacy of quantum SVCs compared to the classical SVC when confronted with limited training data instances.

Table \ref{tab:staf-result} presents the findings of our investigation across eight subject systems, each containing training data instances numbering approximately between 350 and 500. We calculate Precision, recall, F-score, ROC-AUC, and Matthews correlation for each support vector classifier examined in this study. The performance metrics listed in this table illustrate a comparative analysis of support vector classifiers. The highest value for each performance metric is denoted in bold font within each column. Notably, the subject systems Tomcat and Ambari consistently exhibit superior performance across all criteria when compared to SVC and PQSVC. Within the Jackrabbit subject system, QSVC outperforms both algorithms across all performance metrics except Precision, which is also very close to the precision value of SVC. Across the remaining subject systems outlined in the table, QSVC consistently either matches or closely rivals SVC's performance. Conversely, PQSVC consistently performs poorly compared to both SVC and QSVC across all test scenarios.

The concept of STAF aligns well with agile methodologies [\cite{Agile2012}], particularly in scenarios where software is developed and refined through iterative, short sprints. In Agile frameworks, rapid defect detection is critical, especially within the confines of sprint cycles where limited historical data often challenges traditional machine learning models. Our study illustrates that Quantum Support Vector Classifier (QSVC) performs robustly under these conditions, making it a promising candidate for sprint-based defect prediction. The quantum feature map, which encodes data into a high-dimensional quantum space, may provide enhanced pattern recognition for smaller datasets, as observed in Short-term Activity Frames (STAF). This mapping also contributes to QSVC’s superior recall rates in these scenarios compared to classical SVC, which often struggles with limited training instances. The higher recall rate in detecting buggy commits could aid software maintenance teams by identifying potential issues earlier in the development process. By addressing early-stage datasets with relatively fewer instances for software defect prediction, QSVC could be leveraged for its effectiveness in Agile settings, where data accumulation is incremental, and decision-making is time-sensitive. It can also minimize disruption and maintain release schedules during the incremental process without propagating defects of earlier stages to the later stage of a software project.

\begin{table}[]
\centering
\caption{Short-term activity frames (STAF) Result}
\label{tab:staf-result}
\begin{tabular}{|p{1cm}|lccccc|}
\hline
\multicolumn{1}{|c|}{\textbf{Subject Systems}}                  & \multicolumn{1}{c|}{\textbf{Algorithms}} & \multicolumn{1}{c|}{\textbf{Precision}} & \multicolumn{1}{c|}{\textbf{Recall}} & \multicolumn{1}{c|}{\textbf{F1~Score}} & \multicolumn{1}{c|}{\textbf{ROC-AUC}} & \textbf{MCC}  \\ \hline \hline
\multicolumn{1}{|c|}{\multirow{3}{*}{\textbf{Jackrabbit}}} & \multicolumn{1}{l|}{\textbf{SVC}}   & \multicolumn{1}{c|}{\textbf{0.78}}      & \multicolumn{1}{c|}{0.77}            & \multicolumn{1}{c|}{0.77}            & \multicolumn{1}{c|}{0.77}             & 0.55          \\ \cmidrule{2-7} 
\multicolumn{1}{|c|}{}                             & \multicolumn{1}{l|}{\textbf{PQSVC}} & \multicolumn{1}{c|}{0.74}               & \multicolumn{1}{c|}{0.68}            & \multicolumn{1}{c|}{0.71}            & \multicolumn{1}{c|}{0.72}             & 0.44          \\ \cmidrule{2-7} 
\multicolumn{1}{|c|}{}                             & \multicolumn{1}{l|}{\textbf{QSVC}}  & \multicolumn{1}{c|}{0.77}               & \multicolumn{1}{c|}{\textbf{0.88}}   & \multicolumn{1}{c|}{\textbf{0.82}}   & \multicolumn{1}{c|}{\textbf{0.81}}    & \textbf{0.63} \\ \hline \hline

\multicolumn{1}{|c|}{\multirow{3}{*}{\textbf{Bitcoin}}} & \multicolumn{1}{l|}{\textbf{SVC}}   & \multicolumn{1}{c|}{0.84}               & \multicolumn{1}{c|}{0.78}            & \multicolumn{1}{c|}{\textbf{0.81}}   & \multicolumn{1}{c|}{\textbf{0.82}}    & \textbf{0.63} \\ \cmidrule{2-7} 
\multicolumn{1}{|c|}{}                             & \multicolumn{1}{l|}{\textbf{PQSVC}} & \multicolumn{1}{c|}{0.59}               & \multicolumn{1}{c|}{0.58}            & \multicolumn{1}{c|}{0.59}            & \multicolumn{1}{c|}{0.59}             & 0.18          \\ \cmidrule{2-7} 
\multicolumn{1}{|c|}{}                             & \multicolumn{1}{l|}{\textbf{QSVC}}  & \multicolumn{1}{c|}{0.74}               & \multicolumn{1}{c|}{\textbf{0.79}}   & \multicolumn{1}{c|}{0.76}            & \multicolumn{1}{c|}{0.76}             & 0.51          \\ \hline \hline

\multicolumn{1}{|c|}{\multirow{3}{*}{\textbf{QT}}} & \multicolumn{1}{l|}{\textbf{SVC}}   & \multicolumn{1}{c|}{\textbf{0.80}}      & \multicolumn{1}{c|}{0.51}            & \multicolumn{1}{c|}{0.62}            & \multicolumn{1}{c|}{0.69}             & \textbf{0.41} \\ \cmidrule{2-7} 
\multicolumn{1}{|c|}{}                             & \multicolumn{1}{l|}{\textbf{PQSVC}} & \multicolumn{1}{c|}{0.59}               & \multicolumn{1}{c|}{0.36}            & \multicolumn{1}{c|}{0.45}            & \multicolumn{1}{c|}{0.55}             & 0.12          \\ \cmidrule{2-7} 
\multicolumn{1}{|c|}{}                             & \multicolumn{1}{l|}{\textbf{QSVC}}  & \multicolumn{1}{c|}{0.68}               & \multicolumn{1}{c|}{\textbf{0.75}}   & \multicolumn{1}{c|}{\textbf{0.72}}   & \multicolumn{1}{c|}{\textbf{0.70}}    & 0.4           \\ \hline \hline

\multicolumn{1}{|c|}{\multirow{3}{*}{\textbf{Mongo}}} & \multicolumn{1}{l|}{\textbf{SVC}}   & \multicolumn{1}{c|}{\textbf{0.78}}      & \multicolumn{1}{c|}{0.73}            & \multicolumn{1}{c|}{\textbf{0.76}}   & \multicolumn{1}{c|}{\textbf{0.77}}    & \textbf{0.53} \\ \cmidrule{2-7} 
\multicolumn{1}{|c|}{}                             & \multicolumn{1}{l|}{\textbf{PQSVC}} & \multicolumn{1}{c|}{0.55}               & \multicolumn{1}{c|}{0.58}            & \multicolumn{1}{c|}{0.57}            & \multicolumn{1}{c|}{0.56}             & 0.11          \\ \cmidrule{2-7} 
\multicolumn{1}{|c|}{}                             & \multicolumn{1}{l|}{\textbf{QSVC}}  & \multicolumn{1}{c|}{0.70}               & \multicolumn{1}{c|}{\textbf{0.75}}   & \multicolumn{1}{c|}{0.72}            & \multicolumn{1}{c|}{0.72}             & 0.43          \\ \hline \hline

\multicolumn{1}{|c|}{\multirow{3}{*}{\textbf{Oozie}}} & \multicolumn{1}{l|}{\textbf{SVC}}   & \multicolumn{1}{c|}{\textbf{0.79}}      & \multicolumn{1}{c|}{0.76}            & \multicolumn{1}{c|}{\textbf{0.77}}   & \multicolumn{1}{c|}{\textbf{0.78}}    & \textbf{0.55} \\ \cmidrule{2-7} 
\multicolumn{1}{|c|}{}                             & \multicolumn{1}{l|}{\textbf{PQSVC}} & \multicolumn{1}{c|}{0.66}               & \multicolumn{1}{c|}{0.51}            & \multicolumn{1}{c|}{0.58}            & \multicolumn{1}{c|}{0.62}             & 0.25          \\ \cmidrule{2-7} 
\multicolumn{1}{|c|}{}                             & \multicolumn{1}{l|}{\textbf{QSVC}}  & \multicolumn{1}{c|}{0.73}               & \multicolumn{1}{c|}{\textbf{0.77}}   & \multicolumn{1}{c|}{0.75}            & \multicolumn{1}{c|}{0.74}             & 0.49          \\ \hline \hline

\multicolumn{1}{|c|}{\multirow{3}{*}{\textbf{Tomcat}}} & \multicolumn{1}{l|}{\textbf{SVC}}   & \multicolumn{1}{c|}{0.60}               & \multicolumn{1}{c|}{\textbf{0.77}}   & \multicolumn{1}{c|}{0.68}            & \multicolumn{1}{c|}{0.63}             & 0.28          \\ \cmidrule{2-7} 
\multicolumn{1}{|c|}{}                             & \multicolumn{1}{l|}{\textbf{PQSVC}} & \multicolumn{1}{c|}{0.57}               & \multicolumn{1}{c|}{0.51}            & \multicolumn{1}{c|}{0.54}            & \multicolumn{1}{c|}{0.56}             & 0.12          \\ \cmidrule{2-7} 
\multicolumn{1}{|c|}{}                             & \multicolumn{1}{l|}{\textbf{QSVC}}  & \multicolumn{1}{c|}{\textbf{0.66}}      & \multicolumn{1}{c|}{\textbf{0.77}}   & \multicolumn{1}{c|}{\textbf{0.71}}   & \multicolumn{1}{c|}{\textbf{0.69}}    & \textbf{0.38} \\ \hline \hline

\multicolumn{1}{|c|}{\multirow{3}{*}{\textbf{Lucene}}} & \multicolumn{1}{l|}{\textbf{SVC}}   & \multicolumn{1}{c|}{\textbf{0.77}}      & \multicolumn{1}{c|}{0.63}            & \multicolumn{1}{c|}{\textbf{0.69}}   & \multicolumn{1}{c|}{\textbf{0.72}}    & \textbf{0.45} \\ \cmidrule{2-7} 
\multicolumn{1}{|c|}{}                             & \multicolumn{1}{l|}{\textbf{PQSVC}} & \multicolumn{1}{c|}{0.57}               & \multicolumn{1}{c|}{0.45}            & \multicolumn{1}{c|}{0.50}            & \multicolumn{1}{c|}{0.55}             & 0.11          \\ \cmidrule{2-7} 
\multicolumn{1}{|c|}{}                             & \multicolumn{1}{l|}{\textbf{QSVC}}  & \multicolumn{1}{c|}{0.65}               & \multicolumn{1}{c|}{\textbf{0.73}}   & \multicolumn{1}{c|}{\textbf{0.69}}   & \multicolumn{1}{c|}{0.67}             & 0.35          \\ \hline \hline

\multicolumn{1}{|c|}{\multirow{3}{*}{\textbf{Ambari}}} & \multicolumn{1}{l|}{\textbf{SVC}}   & \multicolumn{1}{c|}{\textbf{0.74}}      & \multicolumn{1}{c|}{0.55}            & \multicolumn{1}{c|}{0.63}            & \multicolumn{1}{c|}{0.68}             & 0.37          \\ \cmidrule{2-7} 
\multicolumn{1}{|c|}{}                             & \multicolumn{1}{l|}{\textbf{PQSVC}} & \multicolumn{1}{c|}{0.56}               & \multicolumn{1}{c|}{0.39}            & \multicolumn{1}{c|}{0.46}            & \multicolumn{1}{c|}{0.54}             & 0.08          \\ \cmidrule{2-7} 
\multicolumn{1}{|c|}{}                             & \multicolumn{1}{l|}{\textbf{QSVC}}  & \multicolumn{1}{c|}{\textbf{0.74}}      & \multicolumn{1}{c|}{\textbf{0.88}}   & \multicolumn{1}{c|}{\textbf{0.80}}   & \multicolumn{1}{c|}{\textbf{0.79}}    & \textbf{0.58} \\ \hline
\end{tabular}
\end{table}

Our aggregation strategy, which successfully mitigates the limitations of the QSVC algorithm in handling larger datasets, holds potential for application to other Quantum Machine Learning (QML) algorithms. Given that many QML algorithms face similar scalability challenges due to the exponential runtime demands associated with quantum feature mapping. The aggregation of chunk models offers a pathway to improve their practicality for real-world, large-scale datasets from various software projects. By enabling predictions on smaller data segments and combining the results, this method can extend beyond QSVC, potentially enhancing the applicability of QML models like Quantum Neural Networks \cite {Beer2020} and Quantum k-Nearest Neighbors \cite {Zardini2024, Jing2022} in similar large-data contexts.

\begin{tcolorbox}[colback=gray!10!white, colframe=gray!80!black, arc=4mm]
\textbf{RQ2:} Can we apply Quantum SVC algorithms on a large dataset of real-life software bug detection problems?
\end{tcolorbox}

While SVC and PQSVC algorithms maintain responsiveness and generate results even with datasets exceeding 500 instances within a realistic timeframe, QSVC exhibits non-responsiveness and fails to produce results with datasets of similar size in our investigation. Although the PQSVC algorithm demonstrates the potential to handle larger datasets, it exhibits inferior performance compared to classical SVC. The key interest in our study is the dataset challenge posed by QSVC, which initially showed promise in delivering superior results compared to classical SVC and PQSVC for smaller datasets but failed to produce a result from large datasets due to its runtime requirements. 
Consequently, we investigated whether QSVC could be effectively utilized with larger datasets and compared its performance with classical SVC and PQSVC, to answer the \textbf{RQ2} of this investigation. 

\begin{figure}[htp]
\centering
\includegraphics[width=0.90\textwidth] {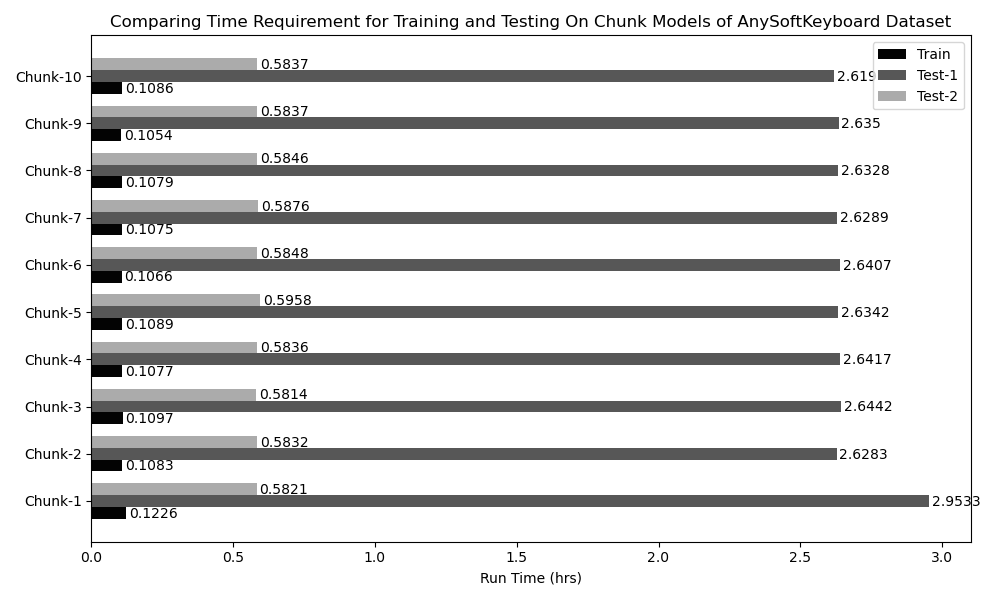}
\caption{Comparison of Training and Testing Runtimes for Chunk Models of the QSVC Algorithm Using the AnySoftK Dataset. Chunk models are trained on 500 data instances and evaluated on 1721 test instances in two test scenarios, Test-1 and Test-2.}
\label{fig:compTrainTest}
\end{figure}

Figure \ref{fig:compTrainTest} illustrates a comparative analysis of the time demands associated with the QSVC algorithm on the largest dataset examined in our study, namely AnySoftK. To effectively manage the scale of this dataset, we segmented the training data into chunks, each comprising 500 instances, and subsequently trained distinct QSVC models on these subsets. Given that the AnySoftK dataset consists of 6883 training instances, this partitioning resulted in 14 chunks, with 13 chunks containing 500 instances each and the final chunk accommodating 383 commit instances. We present a runtime comparison focusing on 10 of these chunks in the figure, as the trends observed in these samples were representative of the other chunks of the dataset. During both the training and testing phases, we noted a consistent pattern of time comparison across various software projects when employing the investigation.

This figure provides insights into the time requirements, measured in hours, for both the training and testing phases. Testing was conducted under two distinct scenarios, denoted as Test-1 and Test-2. In Test-1, we followed the conventional approach, supplying the entire test dataset for detection using any machine learning algorithm. Specifically, this involved executing the code fragment: 
\texttt{classifier.predict(allTestFeatures)}. Acknowledging the significant time requirement in the ``Test-1'' approach, we delved into the incremental testing approach denoted as ``Test-2" where we provide one test instance at a time and iterate the whole process for the entire test instances in the dataset. From the figure, it is visible that, the ``Test-2" approach largely reduces the overall testing time compared to the ``Test-1" approach providing the same test result on the test dataset. Such an incremental testing strategy is also uniquely identified in our investigation and it can speed up the quantum algorithm testing approach to a higher extent.  


The QSVC algorithm, derived from the classical SVC in the Scikit-learn library, operates similarly to its classical counterpart, with the key distinction being the utilization of a quantum kernel for training and testing. Upon receiving a dataset, the algorithm initially transforms and maps it to a quantum feature space, after which the resulting feature map is forwarded to the classical SVC for the \texttt{classifier.fit()} or \texttt{classifier.predict()} operation. Notably, the computational complexity of this feature transformation and mapping process escalates exponentially with the number of data instances. Challenges arise when tackling classification problems with extensive feature spaces, where estimating kernel functions becomes computationally demanding, thus limiting successful solutions [\cite{QuantumEnhancedFS}]. Consequently, to get a successful solution in our investigation during the application of QSVC with a large test set, we explore the impact of testing a reduced number of data instances per iteration to mitigate this complexity. Specifically, we adopt an iterative approach where one data instance is provided in each iteration during the testing phase (referred to as Test-2), and we assess the cumulative runtime requirement for processing the entire test dataset, comprising 1721 data instances. Our investigation, as illustrated in Figure \ref{fig:compTrainTest}, reveals that employing the Test-2 strategy noticeably reduces the overall runtime demands while yielding the same detection result compared to the $Test-1$ strategy.

\begin{tcolorbox}[colback=gray!10!white, colframe=gray!80!black, arc=4mm]
\textbf{RQ3:} Does aggregation of trained QSVC models on smaller chunks of datasets make a better globally trained QSVC model to deal with large datasets?
\end{tcolorbox}

To address research question RQ3, we carefully selected six datasets with sufficient data to construct at least two segments, each comprising approximately 500 data instances, as outlined in Table \ref{tab:dataset-summary}. The methodology employed in this section is visually depicted in Figure \ref{fig:chunkAggregate}. We conducted separate training sessions for QSVC models using each data chunk and preserving the trained models on disk for subsequent use. Following the completion of the training process across all data segments for the six subject systems, we initiated the tuning process to determine the aggregation threshold for the trained chunk models associated with each subject system. Once the aggregation threshold was determined, we executed the aggregation process to derive the detection outcomes for buggy software commits and subsequently reported our findings. Our findings are summarized in Table \ref{tab:aggregate-result}, which offers a comparative analysis of SVC, PQSVC, individual top-3 QSVC chunk models, and Global QSVC performance on the test datasets. Remarkable improvements achieved by these algorithms are highlighted in boldface font.

\begin{longtable}{||p{2cm}|c|c|c|c|c|c|c|}
\caption{Aggregation of Chunk Models to Global QSVC Result Comparison} \label{tab:aggregate-result} \\
\hline
\textbf{Projects} & \rotatebox{90}{\textbf{Algorithms}} & \rotatebox{90}{\textbf{Accuracy}} & \rotatebox{90}{\textbf{Precision}} & \rotatebox{90}{\textbf{Recall}} & \rotatebox{90}{\textbf{F1~Score}} & \rotatebox{90}{\textbf{ROC-AUC}} & \rotatebox{90}{\textbf{MCC}} \\ \hline \hline
\endfirsthead
\caption[]{Aggregation of Chunk Models to Global QSVC Result Comparison (continued)} \\
\hline
\textbf{Projects} & \rotatebox{90}{\textbf{Algorithms}} & \rotatebox{90}{\textbf{Accuracy}} & \rotatebox{90}{\textbf{Precision}} & \rotatebox{90}{\textbf{Recall}} & \rotatebox{90}{\textbf{F1~Score}} & \rotatebox{90}{\textbf{ROC-AUC}} & \rotatebox{90}{\textbf{MCC}} \\ \hline \hline
\endhead

\multirow{6}{*}{\textbf{AnySoftK}} & \textbf{SVC} & \textbf{0.71} & \textbf{0.73} & 0.69 & \textbf{0.71} & \textbf{0.71} & \textbf{0.42} \\ \cmidrule{2-8} 
& \textbf{PQSVC} & 0.57 & 0.61 & 0.46 & 0.52 & 0.57 & 0.15 \\ \cmidrule{2-8} 
& \textbf{QSVC (1)} & 0.64 & 0.64 & 0.71 & 0.67 & 0.64 & 0.29 \\ \cmidrule{2-8} 
& \textbf{QSVC (2)} & 0.64 & 0.64 & 0.69 & 0.66 & 0.64 & 0.29 \\ \cmidrule{2-8} 
& \textbf{QSVC (3)} & 0.64 & 0.64 & 0.68 & 0.66 & 0.64 & 0.28 \\ \cmidrule{2-8} 
& \textbf{G. QSVC} & 0.67 & 0.65 & \textbf{0.76} & 0.70 & 0.67 & 0.34 \\ \hline \hline

\multirow{6}{*}{\textbf{Facebook}} & \textbf{SVC} & 0.76 & \textbf{0.83} & 0.66 & 0.74 & 0.76 & 0.54 \\ \cmidrule{2-8} 
& \textbf{PQSVC} & 0.59 & 0.61 & 0.54 & 0.57 & 0.60 & 0.19 \\ \cmidrule{2-8} 
& \textbf{QSVC (1)} & 0.64 & 0.64 & 0.71 & 0.67 & 0.64 & 0.29 \\ \cmidrule{2-8} 
& \textbf{QSVC (2)} & 0.8 & 0.78 & 0.85 & 0.81 & 0.80 & 0.61 \\ \cmidrule{2-8} 
& \textbf{QSVC (3)} & 0.79 & 0.77 & 0.82 & 0.79 & 0.79 & 0.58 \\ \cmidrule{2-8} 
& \textbf{G. QSVC} & \textbf{0.83} & 0.82 & \textbf{0.86} & \textbf{0.83} & \textbf{0.83} & \textbf{0.66} \\ \hline \hline

\multirow{6}{*}{\textbf{Kiwis}} & \textbf{SVC} & \textbf{0.69} & \textbf{0.72} & 0.56 & 0.63 & 0.68 & \textbf{0.37} \\ \cmidrule{2-8} 
& \textbf{PQSVC} & 0.53 & 0.51 & 0.95 & 0.66 & 0.55 & 0.15 \\ \cmidrule{2-8} 
& \textbf{QSVC (1)} & 0.66 & 0.64 & 0.66 & 0.65 & \textbf{0.66} & 0.31 \\ \cmidrule{2-8} 
& \textbf{QSVC (2)} & 0.64 & 0.62 & 0.66 & 0.64 & 0.64 & 0.28 \\ \cmidrule{2-8} 
& \textbf{QSVC (3)} & 0.64 & 0.62 & 0.66 & 0.64 & 0.64 & 0.28 \\ \cmidrule{2-8} 
& \textbf{G. QSVC} & 0.65 & 0.60 & \textbf{0.84} & \textbf{0.70} & 0.65 & 0.33 \\ \hline \hline

\multirow{6}{*}{\textbf{Jm1}} & \textbf{SVC} & \textbf{0.66} & \textbf{0.73} & 0.54 & 0.62 & \textbf{0.66} & \textbf{0.34} \\ \cmidrule{2-8} 
& \textbf{PQSVC} & 0.50 & 0.51 & 0.97$^*$ & 0.67 & 0.49 & -0.04 \\ \cmidrule{2-8} 
& \textbf{QSVC (1)} & 0.53 & 0.52 & 0.90 & \textbf{0.66} & 0.52 & 0.06 \\ \cmidrule{2-8} 
& \textbf{QSVC (2)} & 0.53 & 0.53 & \textbf{0.73} & 0.62 & 0.53 & 0.06 \\ \cmidrule{2-8} 
& \textbf{QSVC (3)} & 0.56 & 0.56 & 0.68 & 0.61 & 0.56 & 0.12 \\ \cmidrule{2-8} 
& \textbf{G. QSVC} & 0.56 & 0.56 & 0.59 & 0.58 & 0.56 & 0.11 \\ \hline \hline

\pagebreak

\multirow{5}{*}{\textbf{Camel}} & \textbf{SVC} & 0.75 & 0.77 & 0.70 & 0.74 & 0.75 & 0.49 \\ \cmidrule{2-8} 
& \textbf{PQSVC} & 0.65 & 0.65 & 0.64 & 0.64 & 0.65 & 0.30 \\ \cmidrule{2-8} 
& \textbf{QSVC (1)} & 0.80 & 0.76 & \textbf{0.86} & 0.81 & 0.80 & 0.60 \\ \cmidrule{2-8} 
& \textbf{QSVC (2)} & 0.74 & 0.69 & 0.90 & 0.78 & 0.74 & 0.51 \\ \cmidrule{2-8} 
& \textbf{G. QSVC} & \textbf{0.83} & \textbf{0.85} & 0.82 & \textbf{0.83} & \textbf{0.84} & \textbf{0.67} \\ \hline \hline

\multirow{5}{*}{\textbf{OS}} & \textbf{SVC} & 0.37 & 0.32 & 0.24 & 0.28 & 0.37 & -0.27 \\ \cmidrule{2-8} 
& \textbf{PQSVC} & 0.46 & 0.48 & \textbf{0.87} & \textbf{0.62} & 0.46 & -0.13 \\ \cmidrule{2-8} 
& \textbf{QSVC (1)} & 0.60 & 0.62 & 0.51 & 0.56 & 0.60 & 0.20 \\ \cmidrule{2-8} 
& \textbf{QSVC (2)} & 0.55 & 0.55 & 0.57 & 0.56 & 0.55 & 0.10 \\ \cmidrule{2-8} 
& \textbf{G. QSVC} & \textbf{0.65} & \textbf{0.73} & 0.47 & 0.57 & \textbf{0.65} & \textbf{0.32} \\ \hline

\multicolumn{8}{p{0.95\textwidth}}{\footnotesize{* We did not highlight PQSVC in JM1's result as a manual evaluation of the prediction confusion matrix revealed that it did not work with this dataset. In the JM1 dataset, PQSVC classified almost all the test samples as buggy, resulting in a very high recall value. We found that QSVC (Chunk-2) performed best with this dataset.}}
\end{longtable}

\begin{figure}[ht]
\centering
\includegraphics[width=\textwidth] {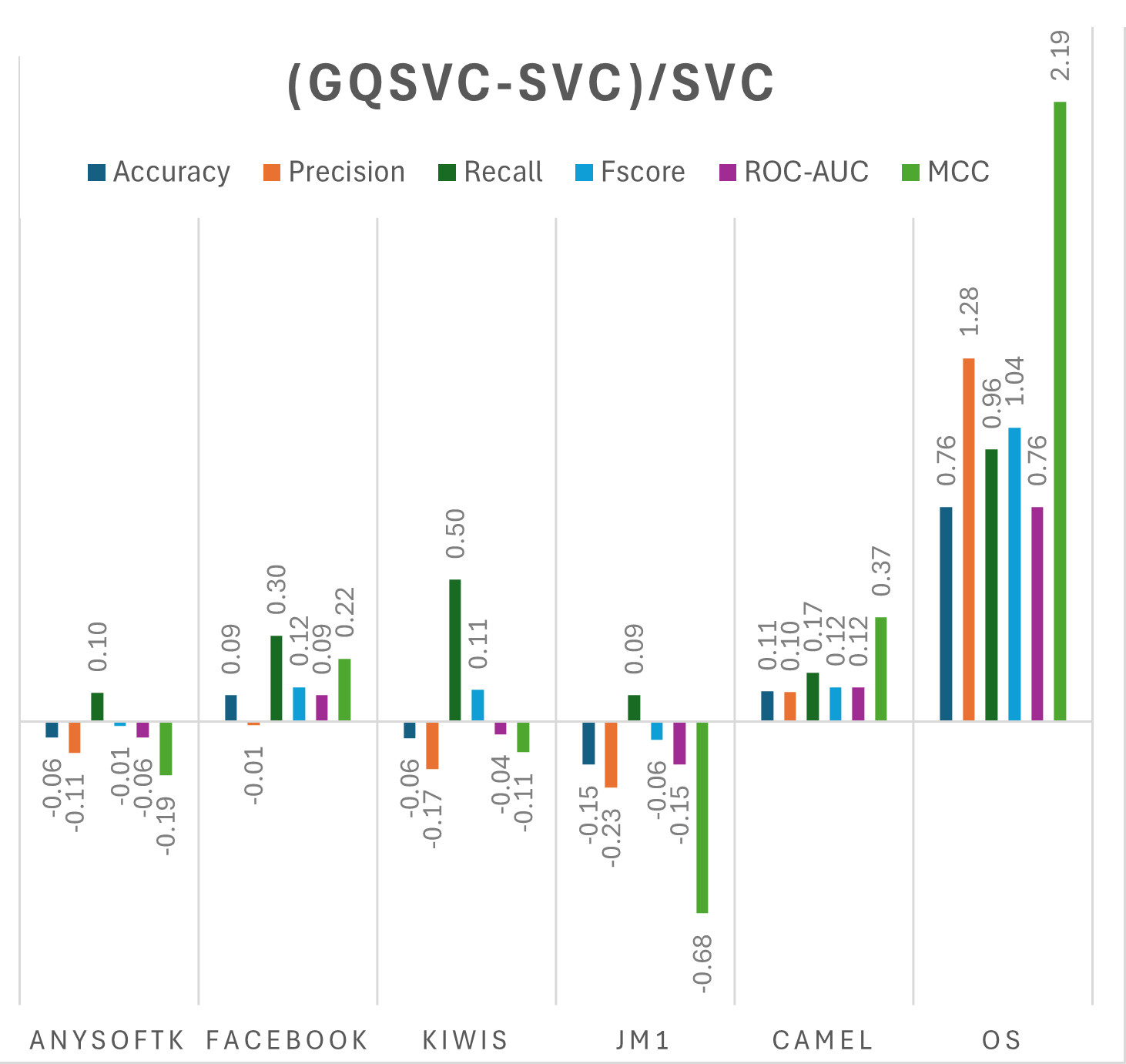}
\caption{Improvements in performance metrics of Global QSVC compared to the classical SVC}
\label{fig:compImprovements}
\end{figure}

Table \ref{tab:aggregate-result} reveals that while individual QSVC chunk models and PQSVC may not perform as effectively as SVC, aggregated Global QSVC demonstrates promising results across various test cases, particularly with the datasets of subject systems like Facebook, Camel, and OpenStack. In three other subject systems, QSVC and Global QSVC exhibit superior performance in terms of recall for buggy commit detection, with minimal compromise on precision. However, it is important to note that despite the notable increase in recall observed in PQSVC (e.g., Jm1 and OpenStack), the negative Mathew correlation (MCC) indicates that these classifiers are not performing well, further confirmed by a manual examination of the classification confusion matrix. This inspection revealed that these classifiers indiscriminately labeled almost all test samples as buggy, resulting in high recall but noticeably compromised precision.

We additionally conduct a comparative analysis presenting the improvements in the performance metric values achieved by the Global QSVC algorithms in contrast to their classical SVC counterparts, as illustrated in Figure \ref{fig:compImprovements}. This visual representation encapsulates and reaffirms our observations across the six subject systems under examination. Notably, big improvements are apparent in the accurate detection of buggy commits within the test datasets of Camel and OpenStack. Within the Facebook test dataset, while there is a marginal 1\% decrease in precision, all other performance metrics exhibit improvement. Although the remaining three subject systems exhibit a reduction in performance metric values, the magnitude of these declines is comparatively minor compared to the enhancements observed in the other subject systems.

\section{Threats to Validity}
\label{threats-validity}
\textbf{Dataset Selection Bias:} The selection of subject systems and datasets might introduce dataset selection bias, impacting the generalizability of findings. We performed this investigation using datasets from 14 software projects; including different datasets might show different findings, potentially limiting the applicability of results to other contexts. To mitigate this concern, we carefully selected datasets with diverse sizes of data instances for analysis. Furthermore, we executed multiple training and testing cycles of classical SVC, PQSVC, and QSVC algorithms. Specifically, we conducted these cycles on eight smaller datasets and more than 60 segments of larger datasets, ensuring a robust evaluation process with plenty of variations. The consistent plot of F-Scores depicted in Figure \ref{fig:validationVsTest} across both tuning and testing phases provides convincing evidence regarding the generalizability of our investigation. In our future studies, we aim to validate our findings of this study with a more extensive array of datasets, encompassing different software domains, development styles, and defect patterns to improve generalizability further.

\textbf{Experimental Design:} The experimental design, including the choice of aggregation thresholds and strategies, may introduce biases or confounding factors. Different choices for segmentation while preparing the dataset chunks or aggregation approaches could yield divergent results and conclusions. To mitigate these concerns, we shuffled each training dataset before taking segments for each data chunk and meticulously determined the aggregation threshold through a tuning strategy, where dataset segments were derived from the training dataset itself. The comparison of performance metrics across tuning and testing phases consistently demonstrates the robustness of our approach, thereby bolstering the validity of the current investigation. Moving forward, our research agenda includes exploring diverse aggregation approaches across Quantum and Classical algorithms to delve deeper into generalizability issues.

\textbf{Scalability of Quantum Algorithms:} The scalability of QSVC algorithm is a significant limitation due to its exponential runtime requirements when processing large datasets. Although our approach of training QSVC models on smaller dataset chunks and combining predictions using aggregation strategies provides a feasible workaround, this method does not fully address the fundamental scalability issues of quantum feature mapping. As such, utility of QSVC in practical large-scale scenarios remains constrained. Future research will explore alternative quantum algorithms and optimizations, such as hybrid quantum-classical methods, to address these scalability challenges more effectively.

\textbf{Evaluation Metrics:} We use the performance metric values, accuracy, precision, recall, F1 Score, ROC-AUC, and MCC to evaluate the findings of our study. However, in some scenarios, these performance metrics might not represent the actual scenario of the investigation, but these are the most widely used metric values in the related studies for software bug prediction [\cite{Kamei2013, Yang2015, Qiu2019, CHEN2023, Zhou2022}].  We showed the relation of these performance metric values of this study to real-world applications. Such as the higher recall of detecting buggy code instances could influence the decision-making process about maintaining code quality and reducing long-term maintenance costs in a regular phase of a software development life cycle.  Specifically, we highlight the importance of recall in high-stakes bug detection for early-stage software releases, where missing bugs could significantly impact the stability of a software project. In the updated discussion, we expand on how these metrics relate to practical software engineering scenarios, such as prioritizing bug detection accuracy during critical development phases. This addition will help readers see the practical relevance of each performance metric. This multifaceted approach mitigates the risk of bias associated with relying solely on specific evaluation metrics and enhances the robustness and validity of our research findings.

\section{Related Work}
\label{related-work}
Quantum computing [\cite{QuantumComputing}], a cutting-edge field, offers tremendous potential for solving complex problems due to its unique features like qubits, superposition, and entanglement. [\cite{QSE2022software}]. However, realizing this potential requires advanced quantum software, which necessitates a dedicated quantum stack comprising operating systems, compilers, and programming languages  [\cite{QProgramming, Qiskit, GroverAlgo, QPCA, lamata2018quantum}]. Developing quantum software poses unique challenges compared to classical software due to quantum computing's inherent characteristics. Addressing these challenges requires innovative approaches in quantum software engineering (QSE) [\cite{QSESerrano}], which involves adapting classical software engineering methodologies to accommodate the probabilistic nature of quantum programs and the complexities of debugging in the presence of intricate quantum states.

In recent years, there has been a notable rise in interest in quantum machine learning, with several studies delving into various aspects of the field [\cite{biamonte2017quantum, Ciliberto2018quantum, Halvicek2019SupervisedQ}]. These studies have introduced new approaches to quantum machine learning algorithms [\cite{biamonte2017quantum}], such as quantum support vector machines (QSVMs) [\cite{Halvicek2019SupervisedQ, QSVC}] and quantum neural networks (QNNs) [\cite{Abbas2021, PRXQuantumLearnability, Jeswal2019QNN, schuld2014quest}]. Notably, quanvolutional neural networks have been proposed as an innovative concept, utilizing quanvolutional layers driven by random quantum circuits to process input data, deviating from traditional convolutional filters [\cite{Henderson2020QuanvolutionalNN}]. Additionally, a unique hybrid method called Quantum Short Long-Term Memory has been suggested, demonstrating the fusion of classical and quantum techniques [\cite{ChenQLSTM}].

Integration of Quantum computing into software engineering is still limited. In a study by \citet{dynamicST}, QML algorithms were examined in dynamic software testing. Our research takes a significant stride toward a similar goal, marking a pioneering effort by conducting a comprehensive analysis, comparing performance, and addressing challenges associated with QML Vs. CML algorithms.

Quantum computing hasn't seen wide adoption in software engineering. \citet{dynamicST} studied how QML algorithms can be applied in dynamic software testing. Meanwhile, \citet{BugInQuantumHuang} presented quantum program assertions for verifying expected quantum states to prevent bugs, showcasing its effectiveness with benchmark programs in various fields like factoring, search, and chemistry. Our research represents a significant advancement in this direction, being one of the first to conduct a thorough analysis. We compare the performance and tackle the challenges between QML and CML algorithms, aiming to enhance understanding and utilization in this field.

\section{Conclusion \& Future Work} 
\label{conclusion-future}
Our research explores quantum machine learning algorithms, with a focus on Quantum Support Vector Classifier (QSVC) algorithms, in detecting buggy software commits. Through an investigation aimed at evaluating the performance and feasibility of QSVC in real-world scenarios, we analyzed its strengths and limitations alongside classical Support Vector Classifier (SVC) algorithms.

Our findings indicate that while QSVC demonstrates potential in the context of short-term activity frames (STAF), its application to large datasets remains a significant challenge due to the exponential runtime requirements of quantum feature mapping. Similarly, our experiments with Quantum Random Forests (QRF) revealed limitations in performance relative to classical machine learning algorithms. To address these issues, we proposed a Homogeneous Ensemble approach that uses an aggregation strategy and threshold-tuning techniques to improve prediction outcomes by training QSVC on smaller data subsets. This strategy provided a viable workaround to QSVC's scalability issues, though it does not fully resolve the inherent challenges posed by quantum feature mapping on large datasets.

Our results suggest that QSVC can offer comparable effectiveness to classical ML algorithms in detecting buggy commits under specific conditions. However, the observed performance margin between quantum and classical methods remains modest. While these findings highlight the potential of quantum machine learning in software engineering, they also underscore the growing state of the field and the need for further exploration to establish the robustness and scalability of these techniques.

Future work will focus on extending our analysis to a broader range of datasets across diverse programming languages and project types. This will help validate the generalizability of our methods and provide deeper insights into the practical applications of quantum machine learning in software quality assurance and bug detection.

\backmatter

\section*{Supplementary information}
To enhance the reproducibility of our research findings, we have made the complete replication package accessible online\footnote{https://figshare.com/s/603102c8be2a9bdcfb74}. The replication package includes all necessary commands to install the Python environment, datasets, and code to replicate the experiments conducted in this study. 
We believe that sharing the replication package contributes to the transparency and openness of our research, fostering a collaborative environment for further exploration and advancement in the field.


\section*{Acknowledgements}
This research is supported in part by the Natural Sciences and Engineering Research Council of Canada (NSERC), and by the industry-stream NSERC CREATE in Software Analytics Research (SOAR). 


\bibliography{sn-bibliography}

\end{document}